\documentclass[10pt]{article} % For LaTeX2e
\usepackage[accepted]{tmlr}
\usepackage{amsmath,amsfonts,bm}

\def\eqref#1{equation~\ref{#1}}
\def\1{\bm{1}}

\DeclareMathAlphabet{\mathsfit}{\encodingdefault}{\sfdefault}{m}{sl}
\SetMathAlphabet{\mathsfit}{bold}{\encodingdefault}{\sfdefault}{bx}{n}

\usepackage{url}
\usepackage{graphicx}
\usepackage{bm}
\usepackage{subfigure}
\usepackage{tabularx}
\usepackage{textcomp}
\usepackage{xcolor}
\usepackage[
colorlinks=true,
linkcolor=teal,
citecolor=gray,
pageanchor=true,
plainpages=false,
pdfpagelabels,
bookmarks,
bookmarksnumbered,
]{hyperref}
\title{Neural-based Compression Scheme for Solar Image Data}

\author{\name Ali Zafari$^{1}$ \email az00004@mix.wvu.edu
    \AND
    \name Atefeh Khoshkhahtinat$^{1}$ \email ak00043@mix.wvu.edu
    \AND
    \name Jeremy A. Grajeda$^{2}$ \email jgra@nmsu.edu
    \AND
    \name Piyush M. Mehta$^{1}$ \email piyush.mehta@mail.wvu.edu
    \AND
    \name Nasser M. Nasrabadi$^{1}$ \email nasser.nasrabadi@mail.wvu.edu
    \AND
    \name Laura E. Boucheron$^{2}$ \email lboucher@nmsu.edu
    \AND
    \name Barbara J. Thompson$^{3}$ \email barbara.j.thompson@nasa.gov
    \AND
    \name Michael S. F. Kirk$^{3}$ \email michael.s.kirk@nasa.gov
    \AND
    \name Daniel E. da Silva$^{3}$ \email daniel.e.dasilva@nasa.gov
    \\\\\\
    \addr $^{1}$West Virginia University, $^{2}$New Mexico State University, $^{3}$NASA Goddard Space Flight Center
    }
    
\newcommand{\angstrom}{\mbox{\normalfont\AA}}

\begin{document}

\maketitle

\begin{abstract}
Studying the solar system and especially the Sun relies on the data gathered
daily from space missions. These missions are data-intensive and compressing this data to make them efficiently transferable to the ground station is a twofold decision to make. Stronger compression methods, by distorting the data, can increase data throughput at the cost of accuracy which could affect scientific analysis of the data. On the other hand, preserving
subtle details in the compressed data
requires a high amount of data to be transferred, reducing the desired gains from compression. In this work, we propose a neural network-based lossy compression method to be used in NASA's data-intensive imagery missions. We chose NASA's Solar Dynamics Observatory (SDO) mission which transmits 1.4 terabytes of data each day
as a proof of concept for the proposed algorithm. In this work, we propose an adversarially trained neural network, equipped with local and non-local attention modules to capture both the local and global structure of the image resulting in a better trade-off in rate-distortion (RD) compared to conventional hand-engineered codecs. The RD variational autoencoder used in this work is jointly trained with a channel-dependent entropy model as a shared prior between the analysis and synthesis transforms to make the entropy coding of the latent code more effective. We also studied how optimizing perceptual losses could help our neural compressor to preserve high-frequency details of the data in the reconstructed compressed image.  Our neural image compression algorithm outperforms currently-in-use and state-of-the-art codecs such as JPEG and JPEG-2000 in terms of the RD performance when compressing extreme-ultraviolet (EUV) data.  As a proof of concept for use of this algorithm in SDO data analysis, we have performed coronal hole (CH) detection using our compressed images, and generated consistent segmentations, even at a compression rate of $\sim0.1$ bits per pixel (compared to 8 bits per pixel on the original data) using EUV data from SDO. 
\end{abstract}

\section{Introduction}
Learning based image compression outperform \citep{yang2022introduction} almost all traditional codecs including JPEG \citep{jpeg} and JPEG-2000 \citep{jpeg2000}. With a basis in convolutional neural networks (CNNs), the performance of said learned compression methods has been improved through various investigations: enhanced entropy model \citep{minnen2018, minnen2020, qian2022entroformer}, learned representation augmentation via attention \citep{cheng2020, zhu2022transformerbased} and incorporating adversarial training for improved perceptual quality of reconstruction \citep{agustsson2019,blau2019,mentzer2020}.
\begin{figure*}[tp]
    \centering
    \includegraphics[width=0.8\linewidth]{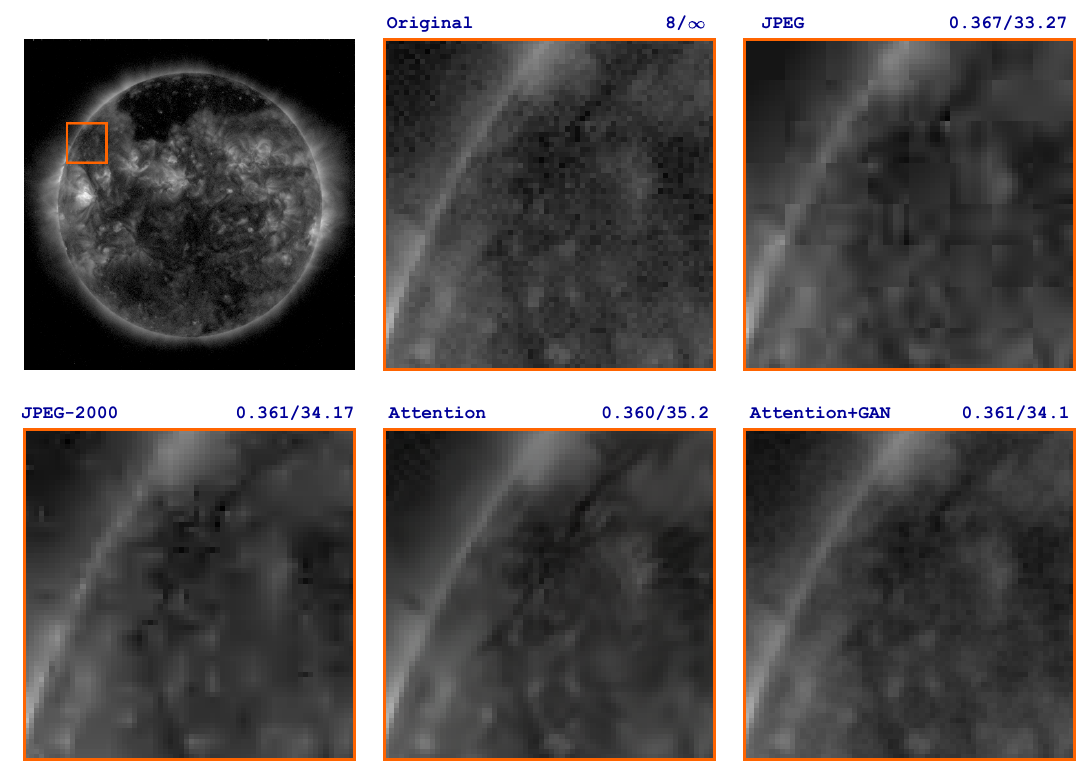}
    \caption{Visual comparison of proposed compression schemes (Attention only and GAN+Attention) to other standard codecs. Reported performance is in terms of bit-rate/distortion [bpp$\downarrow$/PSNR$\uparrow$]. GAN outputs are visually closer to the original input despite their inferior performance in terms of PSNR. \emph{Best viewed on screen.}}
    \label{fig:visual-comparison}
\end{figure*}
In this work, we demonstrate the potential of neural network compression codecs to serve as the go-to approach for future space missions for on-board data compression. We baseline our novel algorithm on the solar image data collected by the Atmospheric Imaging Assembly (AIA) instrument on-board the Solar Dynamics Observatory (SDO) spacecraft and compare the performance with traditional methods. Finally, we show that a CH segmentation~\citep{boucheron2016segmentation} is minimally affected even at extremely low bit-rate compression provided by our proposed neural compression method. Extremely compressed images using our network preserves the required details for the task of CH segmentation on images downloaded from the SDO mission.

\textbf{Contributions of This Work}. Preliminary results of our proposed algorithm is published in \citep{zafari2022compression}, introducing the application of neural image compression on downloaded solar imagery data. The primary contribution of the current work is to critically evaluate the performance of our algorithm on downstream science and/or operational tasks (e.g. CH segmentation). Additionally, we provide a detailed description of the joint forward and backward adapted entropy model used to improve the rate-distortion performance of the proposed algorithm which was not available in \citep{zafari2022compression}.

The paper is structured as follows: Section \ref{sec:related-work} provides a review of neural compression autoencoders and their potential application for a solar mission  with a discussion of the downstream scientific task of segmentation on the SDO data. Section \ref{sec:methods} is devoted to our proposed method. Experimental results and ablation studies are described in Section \ref{sec:experiments}. Section \ref{sec:conclusion} concludes our discussions.

\section{Related Work}\label{sec:related-work}

\subsection{Neural Image Compression}
Transform coding-based image compression algorithms share four main steps to compress an image \citep{goyal2001transformcoding}. 
First, a transform is applied to the image to transfer the pixels from their spatial domain to a transform domain that reduces correlation between pixels and thereby results in many small coefficients (i.e., information packing).
%uncorrelated domain. 
Second, the transformed image is quantized to discard less significant information from the data in the transform domain. Third, entropy coding is utilized to losslessly encode the quantized samples into a stream of ones and zeros. This bitstream is the compressed image. The fourth and final step occurs at the receiving end (or at the reconstructing step), which is responsible for decoding the quantized values to the original space of the input image. The first and most widely used architecture to mimic this scenario in deep neural networks is the convolutional autoencoder which has shown its superiority in terms of rate-distortion performance in the literature \citep{balle2017endtoend}.
Both the encoding and decoding parts of the traditional transform codec can be imitated by an autoencoder \citep{balle2021}.

In an end-to-end optimization of an autoencoder, problems arise when we want to quantize the bottleneck to remove redundancies in order to reach high compression ratios.
ANNs are optimized using gradient descent algorithms which update the parameters of the network by back-propagating the gradients of the loss function. Gradients of the quantization process are not useful for optimizing the loss function as its value is either zero or infinity.
%Thus, all the operations performed in it must be differentiable. 
As a result, we need to approximate hard discrete quantization with an operation yielding informative gradients for updating the parameters of the network. 
The most widely used approach is proposed by \citep{balle2016}, inherited from \citep{gray1998}, in which they showed that adding independent and identically distributed unit uniform noise can be interpreted equivalently as doing scalar quantization on the bottleneck. This method is thoroughly discussed in Section \ref{subsubsec:quantization}. By
%doing so,
applying this noise to the bottleneck, we can optimize the differential entropy of the continuous approximation as a variational upper bound \citep{theis2016} to reduce the entropy of the bottleneck. Low entropy messages are compressed more efficiently into bitstreams \citep{cover1999infotheory}.

 Optimal compression in theory can be achieved by vector quantization \citep{gersho2012vq}. In vector quantization each data point is represented by a prototype and a collection of prototypes, called a codebook, is shared between sender and receiver. The application of vector quantization in ANN-based compression has been investigated by  \citep{agustsson2017}, with the cost of a complicated training procedure. To make the training more accessible, neural compression algorithms follow the classical approach to avoid the complexity of vector quantization. In classical image compression schemes, \emph{e.g.}, JPEG \citep{jpeg}, to get the best out of the quantization process, the first step is to apply an invertible linear transform and translate the image into decorrelated coefficients using a linear transform, \emph{e.g.,} Discrete Cosine Transform (DCT) in JPEG \citep{jpeg}. By doing so, scalar quantization can reach a reasonable performance close to vector quantization \citep{goyal2001transformcoding}. On the other hand, it has been shown \citep{balle2017endtoend, balle2021} that a joint-optimized learned nonlinear transform, \emph{i.e.}, neural network, followed by scalar quantization is sufficient to approximate a parametric form of vector quantization.

As will be shown in Section \ref{sec:methods}, replacing the actual quantization of the latent code/bottleneck of the autoencoder with a uniform noise approximation in the bottleneck of a vanilla autoencoder during training of the network
\citep{balle2018a} will transform it to a Variational Autoencoder (VAE) \citep{kingma2014}.

The difference between the rate-distortion VAE and its vanilla version is the chosen prior for the latent variables. In autoencoder-based image compression, the Gaussian prior of the VAE is replaced with a unit uniform distribution centered on integer numbers to imitate the scalar quantization process.

\begin{figure*}[tp]
    \centering
    \includegraphics[width=\linewidth]{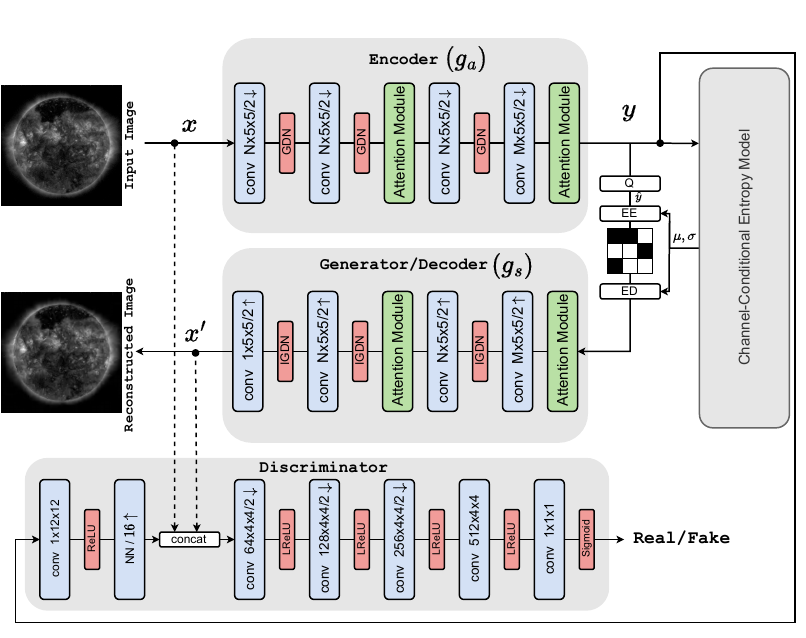}
    \caption{Network architecture. The input image traverses through a series of convolutional layers yielding 16 times smaller (spatially) feature maps than the original input dimensions. To reconstruct the image, the decoder follows the same dimensions of the encoder network in a reverse order using transpose convolution layers.
    A conditional discriminator encourages the generator (decoder) toward better perceptual quality. The number of channels in the encoder and decoder are set as $N=192$ and $M=320$, respectively.  Q performs
    scalar quantization. EE and ED indicate the entropy encoder and decoder, respectively. The checkerboard box represents the bitstream of the compressed image including only zeros and ones. $\mu$ and $\sigma$ are predicted parameters of the latent code probability distribution, defined by the entropy estimation model as shown in Fig. \ref{fig:entropy-model}. LReLU denotes Leaky Rectified Linear Unit activation function. GDN and IGDN correspond to Generalized Divisive Normalization nonlinearity and its inverse, discussed in Section \ref{sec:experiments:implementation}.}
    \label{fig:network-arch}
\end{figure*}

\subsection{Generative Adversarial Training}
Generative Adversarial Networks (GANs) \citep{goodfellow2014} have emerged as a transformative technology in the fields of computer vision and aerospace systems, offering a wide array of innovative applications. In the realm of computer vision, GANs have played a pivotal role in image generation \citep{brock2019biggan, esser2021vqgan}, enhancement \citep{poirier2023restoregan}, and manipulation \citep{ling2021editgan}. Researchers have leveraged GANs to generate high-resolution images from low-resolution inputs \citep{ledig2017srgan}, thereby enhancing the quality of visual data, a crucial advancement for various computer vision applications. GANs have also demonstrated their ability to generate realistic synthetic images \citep{brock2019biggan, esser2021vqgan}, which find application in data augmentation for training deep neural network models, particularly in scenarios where labeled datasets are limited. Additionally, GANs have been instrumental in style transfer \citep{karras2019stylegan}, enabling the creation of unique visual content by merging the style of one image with the content of another \citep{karras2020stylegan2}. GANs are also used in domain translation tasks to make realistic photos out of sketches or colorizing grayscale images \citep{isola2017pixtopix, zhu2017cyclegan}.

The utilization of GANs has been particularly noteworthy in the aerospace field of study. GANs have been employed to enhance satellite imagery for Earth observation and environmental monitoring \citep{jozdani2022ganremotereview}. They facilitate the removal of sensor noise \citep{toner2022noisegan}, the enhancement of image resolution, and the synthesis of missing data \citep{hu2022fieldgan}, or improving the overall quality of satellite-based remote sensing \citep{liu2022ganremote}. GANs has shown great potential in synthesizing RF micro-Doppler signatures to tackle the challenge of limited sample availability and facilitate the training of more complex deep neural networks (DNNs) to improve RF signal classification performance \citep{rahman2022physicsgan}. Training neural networks in a GAN framework also improved unsupervised domain adaptation in satellite pose estimation \citep{wang2023gapgan}. These advancements have the potential to significantly impact industries, from enhancing image analysis to improving the safety and efficiency of aerospace operations.

\subsection{Attention Mechanism in Neural Networks}\label{sec:attention-lit}
When it comes to computer vision, deep convolutional neural networks (CNNs) are the de facto standard despite their poor performance in capturing long-range dependencies \citep{li2021involution,zhou2021decoupled}.
%CNNs will have problems if required to simultaneously capture a few characteristics from non-neighboring pixels.
 The performance degradation of CNNs is attributed to
 %its
 their local receptive field, which is mainly because of the limited kernel size \citep{ramachandran2019stand}
 %in
 of the filters.
%It has been determined that the local nature of kernel sliding on just a few pixels of the input image is the primary cause of this degradation \citep{ramachandran2019stand}.

Efforts have been made to help CNNs capture a more robust representation of the input image. One na\"{i}ve solution is to make the network deeper, but other problems will arise in training such networks which have led to the introduction of deep residual networks (ResNet) \citep{he2016resnet}. Although increasing the parameters of a network will generally lead to a richer representation and better performance, it will make training such networks harder since overfitting can easily happen in these over-parametrized neural networks.
Attention mechanisms have been proposed to address 
%this issue 
the problem of the local receptive field by keeping the depth of the network unchanged. In \citep{wang2017residualattn} a single module was proposed to be included in between sequential convolutional layers, consisting of two branches, the \emph{trunk} to process local features and the \emph{mask} to decide which of the local features in the trunk are more important to be passed to the next convolutional layer, as in Fig. \ref{fig:residual-attention}.

In contrast to local attention, the authors in \citep{wang2018} first discussed how non-local attention can be viewed as a special case of a non-local algorithm which was traditionally used as a method to denoise images \citep{buades2005}. The idea was to find similar pixels/patches in the image/feature map and replace them with a weighted sum over all the others, with higher weights for more similar ones. It can be inferred from  \citep{wang2018} that Vision Transformers (ViT) \citep{dosovitskiy2021vit} are all special cases of the non-local attention mechanism.

The non-local attention block (see Fig. \ref{fig:wnlam}) helps the \emph{mask} branch to efficiently learn the most informative parts of features (in the \emph{trunk}) for the task at hand \citep{zhang2018residualnonlocal}. The authors in \citep{zhang2018residualnonlocal} also added a skip connection to help the output feature maps be richer, by letting the module have access to both attended and raw features. This skip connection prevents vanishing gradients as well. 

Another recently proposed mechanism to incorporate attention in CNNs has been introduced in \citep{woo2018cbam}. It is an enhanced version of the Squeeze-and-Excitation network \citep{hu2018squeeze}, to apply attention on both spatial and channel feature maps separately. This way of applying attention is simpler and computationally more efficient than previous computation-heavy attention mechanisms based on pre-trained networks or complex calculations \citep{xiao2015twolevel}.

\section{Methods}\label{sec:methods}
\begin{figure*}[tp]
    \centering
    \includegraphics[width=0.9\linewidth]{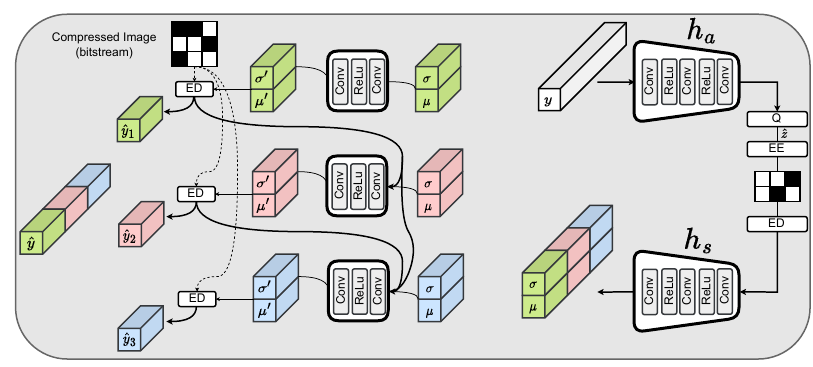}
    \caption{Joint forward and backward adaptation entropy model with the channel conditioning assumption in the probability distribution of the latents. Conv represents a convolutional layer with a kernel size of $3\times3$ with stride and padding of size $1$. Due to limited space, this figure shows channel-wise conditioning on only three slices (red, green, and blue), while in the implementation the latent code is divided into 10 slices. After decoding, the first slice ($\bm{\hat{y}}_1$) is used to decode the second one ($\bm{\hat{y}}_2$), and both of them are used for decoding the third slice ($\bm{\hat{y}}_3$). ReLu stands for a \underline{re}ctified \underline{l}inear \underline{u}nit. Q, EE, and ED denote \underline{q}uantization, and \underline{e}ntropy \underline{e}ncoder/\underline{d}ecoder, respectively. Checkerboard boxes represent bitstream codes. The estimated probability distribution that is parametrized by $\mu$ an $\sigma$ is shared as the prior over the latent code both on the encoder and decoder side at the time of entropy coding.}
    \label{fig:entropy-model}
\end{figure*}
\subsection{Generative Image Compression}
\label{subsec:generative-compression}
Autoencoder-based learned image compression networks, like the one we have proposed in Fig. \ref{fig:network-arch}, generally consist of two major parts. The first part includes the encoder/decoder network and the second part is the bottleneck entropy estimation network. The former is discussed in this section and Section \ref{section:entropy-model} is devoted to describing the functionality of the latter.  As illustrated in Figs. \ref{fig:network-arch} and \ref{fig:entropy-model}, the network input ($\bm{x}$) and output ($\bm{x'}$) relations can be summarized as follows:

\begin{equation}
 \begin{aligned}
    &\bm{x'}=g_s(ED(EE(\bm{\hat{y}}));\bm{\theta_g}),\\
    &\bm{\hat{y}}=\lfloor g_a(\bm{x};\bm{\phi_g})\rceil,\\
    &\bm{\hat{z}}=\lfloor h_a(\bm{y};\bm{\phi_h})\rceil,
\end{aligned}
\end{equation}
in which $\lfloor\cdot\rceil$ denotes the quantization of a real value to the nearest integer number, $\bm{\hat{y}}$ is the quantized latent variable, and $\bm{\hat{z}}$ is its quantized hyper-prior which is discussed in Section \ref{section:entropy-model}.  $ED$ and $EE$ denote the entropy decoder and entropy encoder, respectively. The encoder and decoder nonlinear transforms are represented by $g_a$ and $g_s$ with their learned parameters, $\bm{\phi_g}$ and $\bm{\theta_g}$, respectively. The subscripts $a$ and $s$ refer to \emph{analysis} and \emph{synthesis} as they are common designations for the compression and decompression processes in the area of transform coding-based compression. $h_a$ is the analysis transform to get the hyper-priors of the entropy estimation model, defined by its parameters $\bm{\phi_h}$. Throughout the following section, we use the terms quantized bottleneck and quantized latent code interchangeably.

\subsubsection{Learning Objective} \label{sec:compression:objective}
Any learned image compression network tries to find an
%sweet spot 
optimal tradeoff in the rate-distortion plane by trading off distortion for the expected bitrate (or vice versa), governed by a Lagrangian coefficient $\lambda$.  The rate-distortion trade-off can be described as:
\begin{equation}
    R+\lambda D_r
    \label{eq:rate-distortion},
\end{equation}
where $R$ and $D_r$ correspond to the estimated entropy of the latent code and reconstruction distortion, respectively.
The estimated entropy of the quantized bottleneck, $R$, represents the rate term. Optimizing the parameters of the neural network will enforce this objective to be minimized. The probability distribution of the latent code is variationally approximated by hyper-prior $\bm{z}$. Then the quantized $\bm{\hat{z}}$ is transmitted alongside the compressed image
%known 
as side-information to build the shared prior on the decoder side. Therefore, the entropy of both the quantized bottleneck and its hyper-prior should be optimized:
\begin{equation}
    R=\mathbb{E}_{x\sim p_X}[-\log_2P_{\bm{\hat{y}}|\bm{\hat{z}}}(\bm{\hat{y}}|\bm{\hat{z}};\bm{\theta_h})-\log_2P_{\bm{\hat{z}}}(\bm{\hat{z}};\bm{\psi})],
    \label{eq:rate}
\end{equation}
where $\bm{\theta_h}$ and $\bm{\psi}$ are parameters of the learned entropy model on the latent code ($\bm{\hat{y}}$) and hyper-prior ($\bm{\hat{z}}$), respectively. $P_{\bm{\hat{y}}|\bm{\hat{z}}}$ is the probability mass function of the discrete bottleneck and $P_{\bm{\hat{z}}}$ denotes the probability mass function for the discretized hyper-prior.

In Eq. (\ref{eq:rate-distortion}), $D_r$ accounts for the distortion between the input and output image of the network which can be measured by any desired metric. The prevalently used criterion to measure distortion between input and output is the Mean Squared Error (MSE), which is heavily criticized in the context of computer vision \citep{zhao2017losses} because it often results in the reconstruction of blurry images. Efforts have been made to propose metrics that can adhere perceptually to the human visual system, \emph{e.g.}, the Multi-Scale Structural SIMilarity index (MS-SSIM) \citep{wang2009}. Even these metrics have shown weaknesses when intensely scrutinized \citep{nilsson2020}.
% I'm not sure what "reconstructing blurry images" is, but my primary concern about MSE is that it prioritizes pixel-to-pixel brightness and brighter regions, which cover a small fractional area of the sun.  Regions with less emission do not contribute enough to MSE but most of the solar atmosphere is lower emission.  (BJT) (done/Ali)
%SSIM \citep{wang2004} and MS-SSIM \citep{wang2003}.
% \begin{figure*}[tp]
%     \centering
%     \includegraphics[width=\linewidth]{figs/channelwise.pdf}
%     \caption{{\color{ali}Joint forward and backward adaptation entropy model with the channel conditioning assumption in the probability distribution of the latents. Conv represents a convolutional layer with a kernel size of $3\times3$ with stride and padding of size $1$. Due to limited space, this figure shows channel-wise conditioning on only three slices (red, green, and blue) while in the implementation, the latent code is divided into 10 slices. After decoding, the first slice ($\bm{\hat{y}}_1$) is used to decode the second one ($\bm{\hat{y}}_2$), and both of them are used for decoding the third slice ($\bm{\hat{y}}_3$). ReLu stands for a \underline{re}ctified \underline{l}inear \underline{u}nit. Q, EE, and ED denote \underline{q}uantization, and \underline{e}ntropy \underline{e}ncoder/\underline{d}ecoder, respectively. Checkerboard boxes represent bitstream codes. The estimated probability distribution is shared as the prior over the latent code both on the encoder and decoder side at the time of entropy coding.}}
%     \label{fig:entropy-model}
% \end{figure*}
Recently, perceptual-aware metrics based on features generated by pre-trained neural networks have been proposed. Learned Perceptual Image Patch Similarity (LPIPS) introduced by \citep{zhang2018lpips} uses trained AlexNet/VGGNet features to compare patches of an image with a corresponding reference. In training our neural compressor, we will enhance its reconstruction loss by exploiting this perceptual metric.

To make the reconstruction closer to the input image, we also consider adversarial training for our decoder network. Generative Adversarial Networks (GANs) \citep{goodfellow2014,agustsson2019, mentzer2020hific,blau2019} consisting of a generator and a discriminator sub-network, are able to better match the distribution of data at reconstruction.
%instead of just trying to find the nearest pixel values in order to decrease the distortion. 
In our network, the decoder plays the role of the generator.
%Then
In the GAN framework, the discriminator forces the decoder output to preserve the distribution of the input image at the time of reconstruction. The proposed objective to be optimized for adversarial training of the generator is a combination of distortion and perception as follows:
\begin{equation}
    \begin{split}
    D_r=\mathbb{E}_{\bm{x}\sim p_X}[&\lambda_{recon}MSE(\bm{x},\bm{x'})\\+&\lambda_{perc}LPIPS(\bm{x},\bm{x'})\\-&\lambda_{adv}\log D(\bm{x'},\bm{y})],
    \end{split}
    \label{eq:generator-distortion}
\end{equation}
where $D(\bm{x'},\bm{y})$ denotes the classification decision of the discriminator which is based on two inputs fed into its network, reconstructed image $\bm{x'}$ and bottleneck $\bm{y}$ which serves as a condition. In this setting $\lambda_{adv}$ regularizes how much the discriminator should be optimized to classify the generator reconstructed image (\emph{fake} image in GAN terminology) as an original image. As a result, the generator parameters are optimized with a combination of Eq. (\ref{eq:generator-distortion}) as a distortion and Eq. (\ref{eq:rate}) as a rate penalty.

To make the adversarial training feasible we need the discriminator to judge whether its input sample came from the true distribution of data or is a fake generated image, which is the reconstructed image in our network, \emph{i.e.}, $\bm{x'}$. The discriminator will need to be optimized by a separate auxiliary loss, given as:
\begin{equation}
\begin{split}
    L_{disc.}=&\mathbb{E}_{\bm{x}\sim p_X}[-\log(D(\bm{x},\bm{y})]\\+&\mathbb{E}_{\bm{x}\sim p_X}[-\log(1-D(\bm{x'},\bm{y}))],
\end{split}
\end{equation}
which is the cross entropy loss between the label provided by the discriminator network ($D$) and the true labels, assuming label 1 for the original image and 0 for the reconstructed image by the generator.

It has been shown analytically that increasing the perceptual quality of a generator can result in degradation in terms of distortion measures \citep{blau2018}.
%that the distortion is in direct trade-off with perception \citep{blau2018}. 
GANs are a solution to encourage better perceptual quality in the reconstructed image by tolerating an acceptable amount of distortion. 
%In \citep{blau2019} a third term in this tradeoff was introduced as the rate in the lossy compression scheme. 
More detailed experiments have been adopted in \citep{mentzer2020} to prove in practice the idea that GANs improve perceptual quality at the cost of a small increase in distortion
%increasing the distortion a little bit. 
Therefore, it would be an expected behavior to have a lower Peak Signal to Noise Ratio (PSNR) value on a decoder trained adversarially in contrast to a decoder trained merely on distortion metrics. These adversarially trained networks are expected, however, to perform better when measured with perceptually-motivated metrics.

%\subsubsection{\textbf{Continuous Approximation of Quantization}} \label{sec:methods:compression:quantization}

\subsubsection{Entropy Modeling} \label{section:entropy-model}
The performance of any learned image compression scheme depends heavily on how well it can estimate the true entropy of the bottleneck. Thus the objective will be to minimize the cross entropy between the probability model and the latent code's true probability distribution.
%two.
 To make entropy estimation possible, several probability estimation methods have been proposed in the literature, including empirical histogram density estimation \citep{agustsson2017,theis2017}, piecewise linear models \citep{balle2017endtoend}, conditioning on a latent variable (hyper-prior) \citep{balle2018a}, and context modeling based on autoregressive models \citep{minnen2018}.

From a high-level overview, entropy estimation models can be divided into two main categories: Forward Adaptation (FA) and Backward Adaptation (BA) models. The former suffers from a low capacity to capture all dependencies in the probability distribution of the latent code and the latter's disadvantage is that the decoding process cannot be parallelized. Learned FA models \citep{balle2018a, balle2021} will only use the information provided during the encoding of the image, while BA methods which are based on autoregressive models \citep{minnen2018} need information from the decoded message as well. In the following section, we discuss the functionality of each of them. We emphasize that using a combined entropy model of both FA and BA is the approach we take in this work.  

\paragraph{Forward Adaptation}
To model the probability distribution of latent code $\bm{\hat{y}}$, some assumptions must be made to make the learning feasible and efficient. The simplest way to model any multivariate random variable is to assume independence between all its dimensions, which is called the fully factorized model \citep{bishop2006pattern}, \emph{i.e.}, 
\begin{align}
\begin{split}
    P_{\bm{\hat{y}}}(\bm{\hat{y}})&=P_{\bm{\hat{y}_1}}(\bm{\hat{y}_1})P_{\bm{\hat{y}_2}}(\bm{\hat{y}_2})\dots P_{\bm{\hat{y}_m}}(\bm{\hat{y}_m})\\
    &=\prod_i P_{\bm{\hat{y}_i}}(\bm{\hat{y}_i}),
\end{split}
\end{align}
where $m$ is the dimension of latent code $\bm{\hat{y}}$.

On the other hand, the most flexible and expressive model is to use an autoregressive probability model to capture all dependencies between dimensions of the latent code: 
\begin{align}
\begin{split}
    P_{\bm{\hat{y}}}(\bm{\hat{y}})&=P_{\bm{\hat{y}_1}}(\bm{\hat{y}_1})P_{\bm{\hat{y}_2}}(\bm{\hat{y}_2}|\bm{\hat{y}_{1}})P_{\bm{\hat{y}_3}}(\bm{\hat{y}_3}|\bm{\hat{y}_{1}},\bm{\hat{y}_{2}})\dots\\
    &\dots P_{\bm{\hat{y}_{m-1}}}(\bm{\hat{y}_{m-1}}|\bm{\hat{y}_{<m-1}})P_{\bm{\hat{y}_m}}(\bm{\hat{y}_m}|\bm{\hat{y}_{<m}})\\
    &=\prod_i P_{\bm{\hat{y}_i}}(\bm{\hat{y}_i}|pa(\bm{\hat{y}_i})),
\end{split}
\label{eq:chain-rule}
\end{align}
where $pa(\bm{\hat{y}_i})=\bm{\hat{y}_{<i}}$ denotes the parents of $\bm{\hat{y}_i}$, \emph{i.e.}, $\bm{\hat{y}_1},\bm{\hat{y}_2},\dots,\bm{\hat{y}_{i-1}}$, whose probability density is conditioned on them.
However, letting all the dependencies be visible in the model make it infeasible for practical applications. The curse of dimensionality arises when you model all the conditional dependencies \citep{bishop2006pattern}, preventing the entropy model from being realized. Even if there was enough computational power to learn this fully visible model, the required time to train the model
%have the model trained 
can easily approach infinity as the dimension of the latent code increases. Therefore, an essential decision is how to modify the modeling to capture only essential dependencies while ignoring irrelevant ones.
%modify the dependency modeling to capture only those essential ones and ignore the ones which are not relevant.
%weaken the modeling capability
By doing so the modeling accuracy can be compromised at a reasonable rate but much more efficiently \citep{minnen2018}. 

There is another approach to avoid density modeling based on the probability chain rule Eq. \ref{eq:chain-rule}.
%as shown in Eq. \ref{eq:chain-rule}. 
We can use a latent variable model (LVM) to model the dependencies between visible variables, in our case $\bm{\hat{y}}$, based on hierarchical invisible latent variables \citep{bishop1998latent}, in our case $\bm{\hat{z}}$. By introducing a set of hidden variables, \emph{i.e.}, $\bm{\hat{z}}$, the target random variable, \emph{i.e.}, $\bm{\hat{y}}$, probabilities will be conditionally independent by definition \citep{bishop1998latent}. This is a crucial improvement toward simplifying the modeling complexity and not sacrificing modeling accuracy at the same time. Since, in learned image compression, we are interested in modeling a shared prior to be used both on the encoder and decoder side, this model is called a hyper-prior in the literature \citep{balle2018a,balle2021,qian2021globref,qian2022entroformer,kim2022joint}. Thus, if we denote the hyper-prior by $\bm{\hat{z}}$, the shared prior distribution between the encoder and decoder can be written as:
\begin{align}
    P_{\bm{\hat{y}}}(\bm{\hat{y}})
    &=\prod_i P_{\bm{\hat{y}_i}}(\bm{\hat{y}_i}|\bm{\hat{z}}).
    \label{eq:lvm}
\end{align}
Equation (\ref{eq:lvm}) explicitly models the multivariate random variable by the conditional independence on the hyper-prior $\bm{\hat{z}}$.

\paragraph{Backward Adaptation}
Although ideally latent variable models are able to capture all dependencies in the dimensions of a random variable, the practical issues of training them and the variational and amortization gaps hinder them from performing on par with their autoregressive counterparts.
The variational gap is the mismatch between the assumed variational density and the true distribution of the latent code. The amortization gap refers to the assumption that the posterior is calculated with only a single input to the encoder network.
To address this issue, context should be introduced to the LVM. In addition, to prevent the infeasibility of a global context model (%fully visible 
autoregressive model) the amount of context will be enforced on neighboring elements close to the dimension whose probability is being modeled.
%area near the targeted dimension of the random variable. 
In the area of computer vision, masked convolutions are the de-facto choice to model the local context in a causal manner \citep{van2016pixelcnn, minnen2018}.

\paragraph{Joint Forward and Backward Adaptation}
 To take advantage of both %these models 
LVM, which is an implementation of FA, and autoregressive entropy models which implements the BA modeling  \citep{balle2018a,lee2018contextadaptive,minnen2018,minnen2020}, we define the conditional probability of the latent code as:
\begin{equation}
P_{\bm{\hat{y}}|\bm{\hat{z}}}(\bm{\hat{y}}|\bm{\hat{z}})=\prod_i P(\bm{\hat{y}_i}|\bm{\hat{y}_{j<i}},\bm{\hat{z}};\bm{\theta_h}).
\end{equation}
Conditioning on the quantized hyper-prior, \emph{i.e.}, $\bm{\hat{z}}$, as side-information is an example of FA and conditioning on all previously decoded elements of the latent code, \emph{i.e.}, $\bm{\hat{y}_{j<i}}$, is an example of BA.
%BA performance has been improved in \citep{minnen2020} by letting the conditioning exist only between slices of channels in the bottleneck. 
Spatially autoregressive models have slow decoding time \citep{minnen2018,lee2018contextadaptive} since the decoding time complexity increases quadratically with the spatial dimensions of the latent code. In contrast to spatial autoregressive modeling, \citep{minnen2020} only considers the conditioning of the probabilities on the channels. With this channel-wise autoregressive modeling, the decoding time is only a function of the number of slices assumed in the entropy model thus the spatial dimension of the input image will not affect the decoding latency of our neural image compression network. 
%and it showed that by doing this the decoding process could be reasonably parallelized.
We have used the same approach as in \citep{minnen2020} 
 , as shown in Fig. \ref{fig:entropy-model}, to estimate the entropy and minimize it during
 %the 
 training.

\subsubsection{Relaxed Quantization} \label{subsubsec:quantization}
%The coarse scalar quantization must be replaced by a soft continuous equivalent to guarantee informative gradients
The gradients of uniform scalar quantization are either zero or infinity. Thus this operation is required to be replaced with an approximation that provides informative gradients.
These useful gradients will provide the ability to update the parameters of analysis and synthesis transforms using back-propagation \citep{agustsson2017, balle2016,yang2020bayesquantize,guo2021quantize}.

This approximation can be shown by assuming the simplest form of scalar quantization, \emph{i.e.}, rounding to the nearest integer. If an element of a latent code and its quantized version are denoted by $y$ and $\hat{y}$, respectively, then we have the following:
\begin{align}
    y\sim p_y(y) \quad \xrightarrow{\hspace{0.5cm}\hat{y}=\lfloor y \rceil\hspace{0.5cm}} \quad \hat{y}\sim P_{\hat{y}}(\hat{y}).
\end{align}
%Note that in this setting $y$ and $\hat{y}$ are both continuous random variables. 
In this setting $y$ is a continuous random variable. Although the quantized latent code $\hat{y}$ is a discrete random variable, its generalized probability density function can be written as a train of
%Delta-Dirac 
Dirac delta functions with weights $P_{\hat{y}}(\hat{y}=n)$ at integer-values of $n$ ($n\in \mathbb{Z}$):
\begin{align}
    p_{\hat{y}}(\hat{y})=\sum_{-\infty}^{+\infty}P_{\hat{y}}(n)\delta(\hat{y}-n),
\end{align}
where the weights $P_{\hat{y}}(n)$ define the probability mass function for the discrete random variable $\hat{y}$.

For every integer-valued $\hat{y}$, its probability after being quantized with
%the 
uniform scalar quantization will be:
\begin{equation}
    \begin{aligned}
        P_{\hat{y}}(\hat{y}=n)&=P_{y}(n-\frac{1}{2}<y<n+\frac{1}{2})\\
        &=\int_{n-\frac{1}{2}}^{n+\frac{1}{2}}p_y(\alpha)d\alpha\\
        &=\int_{-\infty}^{+\infty}p_y(\alpha)rect(n-\alpha)d\alpha\\
        &=(p_y*rect)(n),
    \end{aligned}
    \label{eq:convolve}
\end{equation}
where $*$ denotes the convolution operation. Convolution of probability density functions for two independent random variables implies the summation of those random variables.
% We know from the transformation of random variables in probability theory that if two independent random variables are added together, the resulting density function is the convolution of their individual density functions.

From Eq. (\ref{eq:convolve}) we can see that if a unit uniform noise with mean zero, \emph{i.e.}, $\mathcal{U}(-\frac{1}{2}, +\frac{1}{2})$, is added to the unquantized latent representation, it will have the same density value at integer points which are the actual quantized values. Therefore, adding independent zero mean unit uniform noise will act as continuous approximation to the hard uniform scalar quantization.

As a result, we have the relaxed quantized latent code $\Tilde{y}$ as:
\begin{equation}
% \begin{align*}
    \Tilde{y}=y+w,
% \end{align*}
\end{equation}
where $w\sim \mathcal{U}(-\frac{1}{2}, +\frac{1}{2})$.

 We emphasize again that only at integer-valued points (quantized values) the value of the probability density of relaxed latent $\tilde{y}$ will be equal to the probability mass of the actual quantized value $\hat{y}$:
 \begin{equation}
 % \begin{align*}
    P_{\hat{y}}(\hat{y}=n) = p_{\Tilde{y}}(\Tilde{y}=n).   
 % \end{align*}
\end{equation}
\begin{figure*}[tp]
    \centering
    \subfigure[Attention module with skip connection. RB denotes a residual block. WCBAM denotes a window-based convolutional block attention module and WNLAM is a window-based non-local attention module.]{\includegraphics[width=0.49\textwidth]{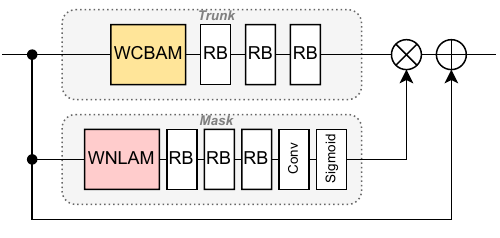} \label{fig:residual-attention}}
    \hfill
    \subfigure[Window-based non-local attention module (WNLAM).]{\includegraphics[width=0.49\textwidth]{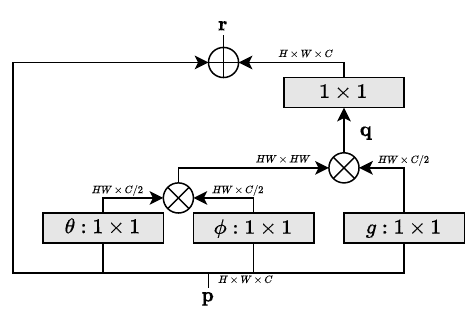}\label{fig:wnlam}}
    \vfill
    \subfigure[Window-based convolutional block attention module (WCBAM). A feature map of C channels with spatial dimensions H$\times$W is the input to the WCBAM block. $w$ is the window size over which to calculate channel attention.]{\includegraphics[width=\textwidth]{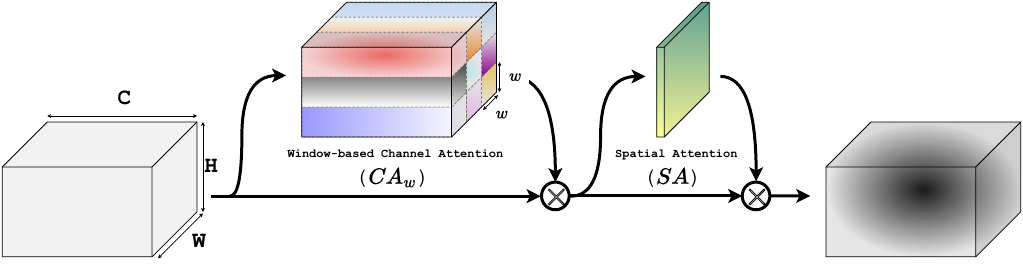} \label{fig:wcbam}}
    \caption{Attention module architecture.}
\end{figure*}
\subsection{Attention Assisted Image Compression}
\label{subsec:attention-asststed}

The attention mechanism in neural networks (discussed in Section \ref{sec:attention-lit}) have been also employed in deep neural compression networks.
\citep{zhou2019attention} applied residual attention, then \citep{chen2021} improved their work by adding a non-local attention mechanism to the mask of the residual attention. As a further improvement, \citep{zou2022} applied non-local attention limited to small windows of the feature maps. This window-based attention attained better results in compression.
Here we propose to use two kinds of attention mechanisms in a window-based manner.

Solar images have a great amount of spatial redundancy compared to natural scene images. In this view, discarding the redundancy and only keeping the low frequency content, which is desired in natural image compression, could lead to high distortions unless paying the cost of transmitting high frequency contents as well. Another important issue when it comes to the compression of solar images is that the minute details and high frequency components are important for the analysis of data (for example solar flare detection and coronal hole segmentation), while in general image compression these high frequency details could be discarded without intolerable cost. To address these differences in the compression domain, we propose two separate attention mechanisms: First, to to apply a window-based non-local attention and second to refine features over a local window in the channel dimension enriching the latent code of the image.

\subsubsection{Window-based Non-Local Attention Module (WNLAM)}
To clarify the procedure of the window-based non-local attention mechanism, described in Section \ref{sec:attention-lit},  a concise review of how this method enriches the representations learned by the convolutional neural networks is included in this section.
A non-local attention block as shown in Fig. \ref{fig:wnlam} is composed of a weighted average, denoted by $\mathbf{q}$,
%(weights calculated as a softmax) 
over a linear transformed version of the block input $\mathbf{p}$, \emph{i.e.}, $g(\mathbf{p})$:
\begin{equation}
    \mathbf{q}_i = \frac{1}{\sum_{\forall j}e^{\theta(\mathbf{p}_i)^T\phi(\mathbf{p}_j)}}\sum_{\forall k}e^{\theta(\mathbf{p}_i)^T\phi(\mathbf{p}_k)}g(\mathbf{p}_k),
    \label{eq:non-local}
\end{equation}
where $g(\cdot)$ is a linear transformation, with learnable parameters $W_g$ implemented by a $1\times1$ convolution layer defined as $g(\mathbf{p_k})=W_g\mathbf{p_k}$. The weights of the sum in Eq. (\ref{eq:non-local}) are calculated by the measure of similarity in the embedding space of the input, \emph{i.e.}, $\theta(\mathbf{p}_i)=W_\theta\mathbf{p_i}$ and $\phi(\mathbf{p}_k)=W_\phi\mathbf{p_k}$, where $W_\theta$ and $W_\phi$ are learnable parameters.

As the final operation in non-local attention, $\mathbf{r}_i$ is calculated by a linear transformation  ($W_r$) added to the original $\mathbf{p}_i$ as follows:
\begin{equation}
    \mathbf{r}_i = W_r\mathbf{q}_i+\mathbf{p}_i.
\end{equation}
Applying a non-local attention mechanism locally through non-overlapping windows has shown to be more effective in the task of image compression \citep{zou2022} than its global counterpart \citep{cheng2020}. In high-bitrate image compression, restoring edges and high-frequency content is as important as representing the global features in the latent representation \citep{jpeg, balle2018a, cheng2020}. Consequently, a na\"{i}ve non-local attention mechanism can perform worse than local attentions which are able to capture local redundancies and preserve details on the reconstructed image \citep{zou2022}.

\subsubsection{Window-based Convolutional Block Attention Module (WCBAM)}
A simple to implement
%kind of 
attention mechanism in CNNs is the convolutional block attention module (CBAM) which has shown great benefit in classification tasks \citep{woo2018cbam}. It includes two attention mechanisms. First, the channel attention ($CA$) guides the network to only consider channels with higher importance for the desired task. Second, the spatial attention ($SA$) dictates where the network should pay more attention. Here we propose to utilize this attention module in a window-based manner. Instead of globally considering the whole spatial extent
%dimensions
of each channel, we focus only on a cropped window size of $w$, as shown in Fig. \ref{fig:wcbam}.

Applying the WCBAM mechanism on the input features $X$ can be summarized as:
\begin{equation}
 \begin{aligned}
    X_{CA} &= CA_w \odot X,\\
    X_{CA,SA} &= SA \odot X_{CA},
\end{aligned}
\label{eq:cbam}
\end{equation}
where $\odot$ is the Hadamard product. $CA_w$ reweighs the channels over each window. After refining the channels, the spatial attention enforces each of the refined channels ($X_{CA}$) to highlight their important spatial content for the task of image compression by the Hadamard product with $SA$.
%denoted by  $\odot$ in Eq. (\ref{eq:cbam}).
%Then $SA$ is multiplied on each refined channel to highlight the important spatial content.

The window-based channel attention is calculated by passing the average and max pool through a shared fully connected network ($F$), as in Eq. (\ref{eq:CAw}):
\begin{equation}
    CA_w = sigmoid(F(Avg(X_w))+F(Max(X_w))),
    \label{eq:CAw}
\end{equation}
where $X_w$ is a chosen window over the input feature map $X$.
Next, the spatial attention weights ($SA$) will be derived by concatenating the average and max pool passed through a convolutional layer as:
\begin{equation}
    SA = sigmoid(Conv([Avg(X_{CA}),Max(X_{CA})])).
\end{equation}
WCBAM  helps the network to capture global dependencies by looking over all channels of each chosen window simultaneously and highlighting the spatially important features with a global average/max pooled feature. These global features are needed in transforming the image from pixel space to feature space.
% , especially those which the window-based non-local attention is incapable of.

\subsubsection{Transformers as Attention Modules}The superiority of models based on transformers, which are a special kind of non-local attention mechanism, compared to convolutional neural networks has been recently proven \citep{dosovitskiy2021vit, liu2021}. Although transformers have shown great benefit in image classification and object detection tasks, their na\"{i}ve application in image compression networks has failed \citep{zou2022}. The goal of transformers is to capture long-range dependencies in an image as opposed to convolutional-based neural networks which inherently have a local inductive bias due to the use of a local kernel. On the other hand, the ultimate goal of image compression is to capture both local and global dependencies in order to summarize them efficiently in the latent code. If the latent code includes global information, we can expect a more compact representation. By na\"{i}vely applying the transformer blocks in the neural compression networks, it was shown \citep{zou2022,zhu2022transformerbased} that optimizing the rate-distortion loss leads to a local receptive field, hindering the self-attention from global dependency modeling. Therefore enriching the attention mechanism in the convolutional neural networks, as we show in this work, could lead to better performance in terms of rate-distortion.

\section{Experiments}\label{sec:experiments}
\begin{figure*}[tp]
    \centering
\subfigure{\includegraphics[width=0.49\textwidth]{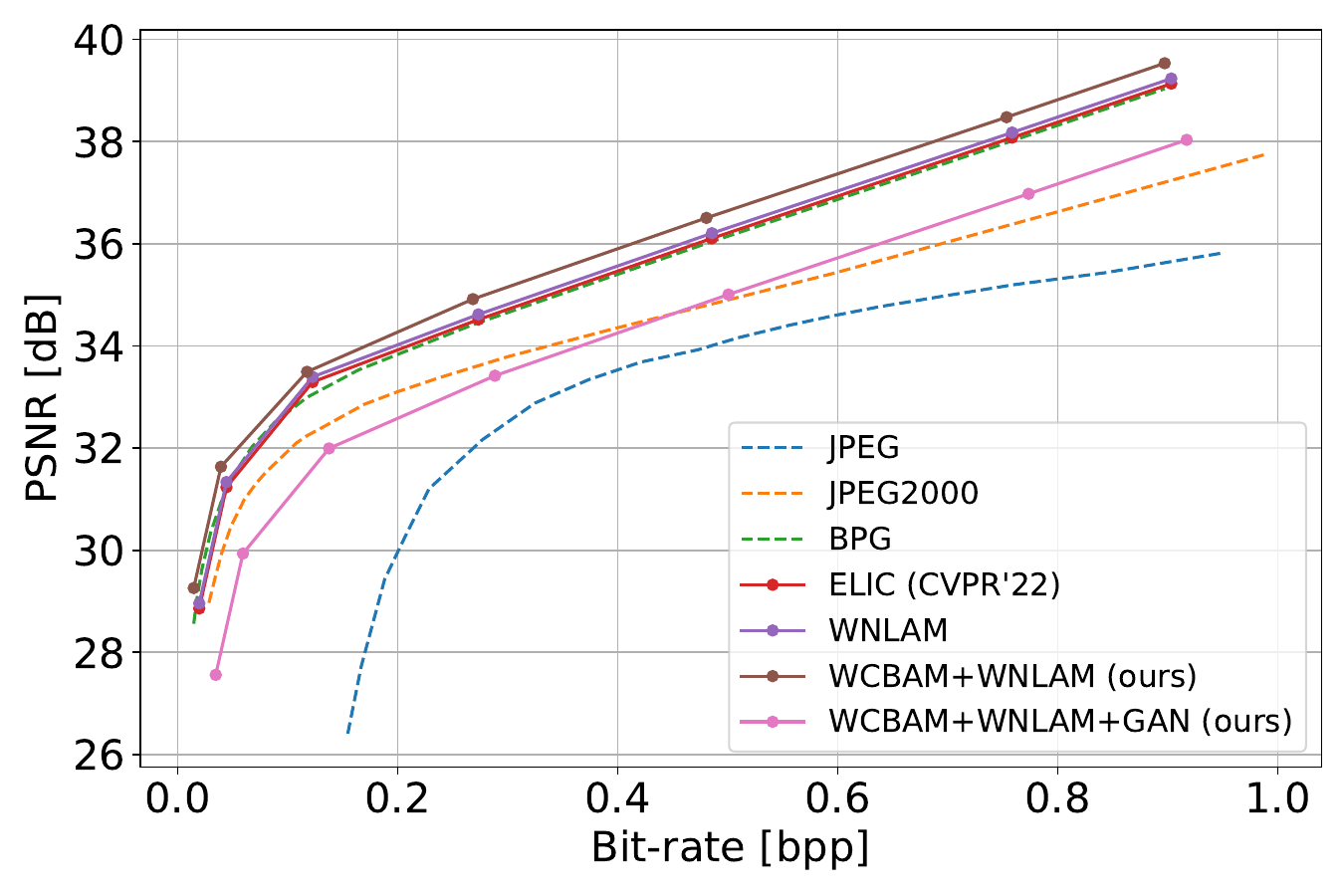}}
    \hfill
    \subfigure{\includegraphics[width=0.49\textwidth]{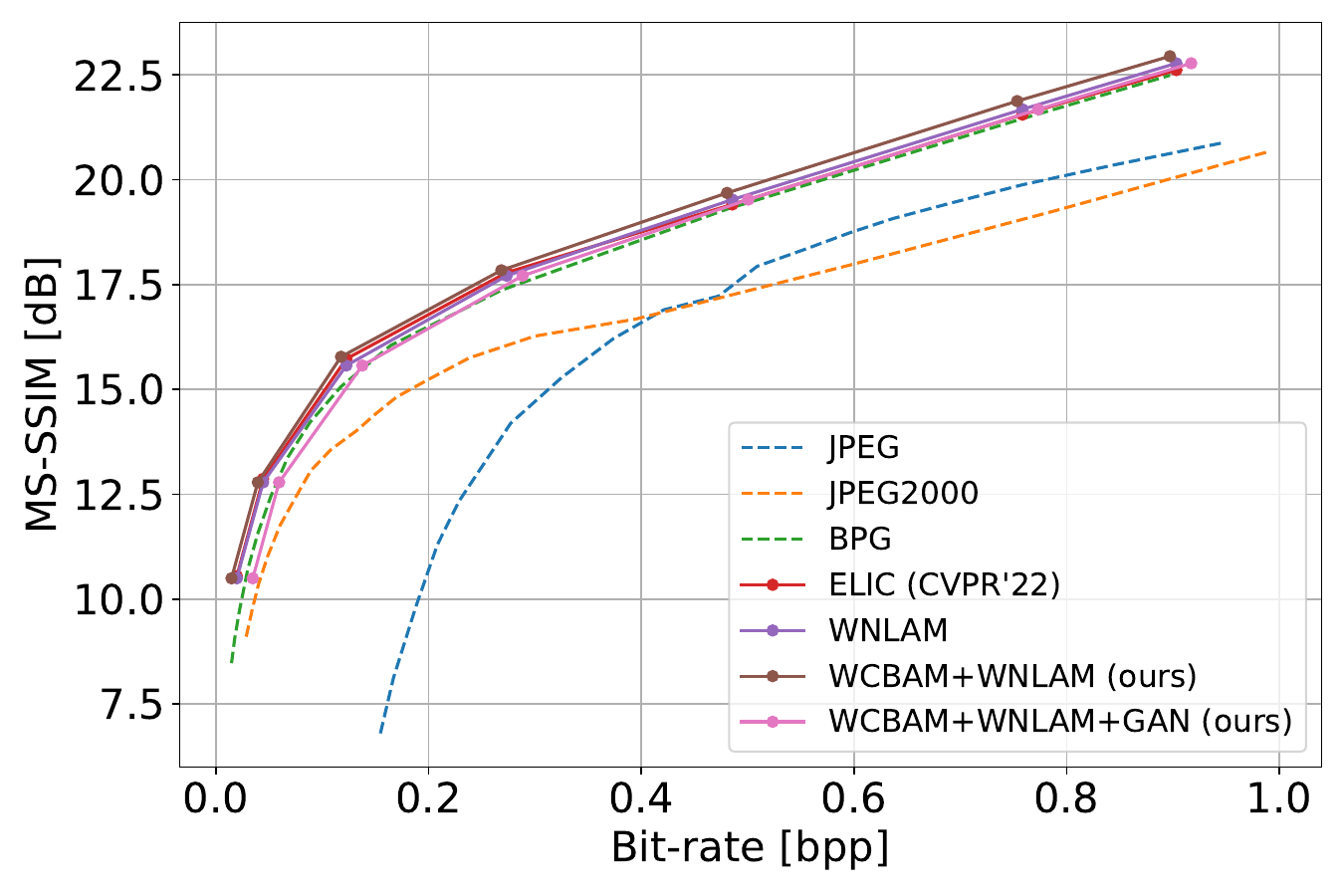}}
    \caption{Rate-distortion curves averaged over the test set described in Section \ref{sec:experiments:dataset}. On the left, PSNR is calculated from MSE using $10\log_{10}\frac{255^2}{MSE}$. On the right, MS-SSIM is reported in logarithmic scale by $-10\log(1-m)$ to show the differences better, in which $m$ is the MS-SSIM in the range of zero to one.}\label{fig:rd-cvrves}
    
\end{figure*}
\subsection{Dataset} \label{sec:experiments:dataset}
The dataset of SDO images described in \citep{galvez2019} includes images of the sun at wavelengths of 94, 131, 171, 193, 211, 304, 335, 1600, and 1700 \angstrom\/ at a cadence of 6 minutes. We temporally downsampled the images to a cadence of 1 hour to decrease dependencies between training samples.
% In addition, to prevent biases of the images with respect to solar cycles, we followed the same approach proposed by \citep{salvatelli2019} to divide the dataset based on the month they are taken. Images of January to August are chosen for training and September to December are reserved for testing. The results reported in this section are all based on this portion of dataset.
In addition, to prevent biases of the images with respect to solar variations at different stages of the solar cycle, we followed the same approach proposed by \citep{salvatelli2019} to divide the dataset based on the month they are taken. Images from January to August of years 2015 to 2018 are chosen for training and September to December of the same years are reserved for testing.
The total number of training images is 21,416 and the test set includes 8,257 samples. The results reported in this section are all based on the test 
portion of the dataset.

% A solar cycle is approximately 11 years, so it should say "to prevent biases of the images with respect to solar variations at different stages of the solar cycle..."  (BJT) (done/Ali)
\subsection{Implementation Details} \label{sec:experiments:implementation}
As the nonlinearity in our neural network, we have utilized a computationally efficient \citep{johnston2019gdnfast} version of Generalized Divisive Normalization (GDN) \citep{balle2016gdn}. As a result of GDN's local normalization, statistical dependencies are reduced in the feature maps. By exploiting GDN instead of more conventional nonlinearities like ReLU, the statistical dependencies in the feature maps will be reduced significantly \citep{balle2016gdn}. 

%Ideally, scalar quantization of a set of decorrelated features will present compression performance close to parametric vector quantization \citep{balle2021}.
During the evaluation phase, entropy coding of the latent integer values was realized by range asymmetric numeral systems coding \citep{duda2013asymmetric}. It is worth mentioning that the entropy coding is lossless and doing it during the training phase has no impact on the measured performance or functionality of the algorithm. It is only during the evaluation phase that entropy coding is needed since the performance of algorithms is compared with standard
%hand-crafted
codecs, such as JPEG \citep{jpeg}, JPEG-2000 \citep{jpeg2000}.
%, and etc.
% \{0.0005, 0.0035, 0.007, 0.009, 0.011, 0.0125, 0.019, 0.025\}
Seven models have been trained with empirically chosen hyper-parameter $\lambda\in\displaystyle\{0.0015, 0.0035, 0.0070, 0.0125, 0.0250, 0.0410, 0.0550\}$ governing the rate-distortion trade-off as in Eq. (\ref{eq:rate-distortion}) for 100 epochs.
%to train each model.
We have used the Adam \citep{kingma15adam} optimizer on batches of size 16 consisting of randomly cropped $256\times256$ patches from the original $512\times512$ images. The initial value of the learning rate is set to $10^{-4}$ and annealed during the training to $1.2\times10^{-6}$  which took about 48 hours on a machine with a single NVIDIA RTX 6000A graphic card, for each model. In addition, to compare our proposed neural codec with state of the art neural-based codecs we picked ELIC \citep{he2022elic} neural compression and trained with exactly the same hyper-parameters and training policies as described for our own networks.
\begin{table*}
\caption{Encoding/decoding latency of the proposed neural-based codec compared other codecs in three different bitrate regimes (input image size of $4096\times4096$ pixels). ENC and DEC refer to encoding and decoding times, respectively and are reported in milliseconds.}\label{table:latency}
\centering
% \tablefont
    \begin{tabularx}{0.95\textwidth}{c|cXXc|cXXc|cXX}
        && \multicolumn{2}{c}{\textit{$\sim 0.1$ bpp}} &&& \multicolumn{2}{c}{\textit{$\sim 0.35$ bpp}}&&&  \multicolumn{2}{c}{\textit{$\sim 0.7$ bpp}}\\\hline

        Codec&&ENC (ms)&DEC (ms)&&&ENC (ms)&DEC (ms)&&&ENC (ms)&DEC (ms) \\\hline
        
        JPEG \citep{jpeg} && 42 & 48 &&&47 & 58 &&& 50 & 64 \\\hline
        
        JPEG2000 \citep{jpeg2000} && 312 & 187 &&& 367 & 221 &&& 416 & 249 \\\hline
        
        BPG \citep{bpg} && 2340 & 2048 &&& 3032 & 2354 &&& 4080 & 2936 \\\hline
        
        ELIC \citep{he2022elic} && 3671 & 3012 &&& 3754 & 3099 &&& 3818 & 3176 \\\hline
        \textbf{ours} && \textbf{3527} & \textbf{3321} &&& \textbf{3654} & \textbf{3423} &&& \textbf{3698} & \textbf{3455} \\
    \end{tabularx}
\end{table*}

Training our autoencoder based on MSE and LPIPS will result in  outperforming even the state-of-the-art hand-engineered codec, \emph{i.e.}, BPG \citep{bpg}, as shown in Fig. \ref{fig:rd-cvrves}. As can be seen in this figure, augmenting the WNLAM attention with WCBAM is capable of improving the rate-distortion performance in terms of both distortion measures PSNR and MS-SSIM. Although the adversarially trained network with the GAN has a distortion performance almost the same as JPEG-2000 \citep{jpeg2000}, the general lower performance of our GAN network is a common issue addressed in \citep{blau2018}. The PSNR or MS-SSIM are unable to capture the perceptual quality of the generated image in a GAN. The perceptual quality of GAN network reconstructions is discussed in Section \ref{sec:experiments:ablation}, where the quality is measured by the perceptual metric LPIPS.

To evaluate our model's computational complexity compared with other compression algorithms, we have conducted experiments to measure the encoding and decoding latencies as reported in Table \ref{table:latency}. All hand-crafted codecs, including JPEG, JPEG2000, and BPG are evaluated on a system powered by Intel Core i7-5930K Broadwell-E CPU. Our proposed method and other state-of-the-art neural compression alogrithm (i.e., ELIC) are tested on a single NVIDIA RTX 6000A graphic card.

\subsection{Ablation Study} \label{sec:experiments:ablation}
To investigate how much the attention modules contribute to the performance of our neural compressor, we have trained three separate networks including a network with only the WNLAM module, then augmented with the WCBAM module, and finally augmented with the WCBAM module and trained adversarially in a GAN framework.
%as the last one, a network with both WNLAM and WCBAM which are trained adversarially in a GAN framework. 
Performance for each of the seven targeted bit-rates is discussed in Section \ref{sec:experiments:implementation}. The first architecture has only the WNLAM module (Fig. \ref{fig:rd-cvrves}) whose performance in terms of PSNR and MS-SSIM has been improved by adding the WCBAM attention module.

\begin{figure}[tp]
    \centering
    \includegraphics[width=0.5\linewidth]{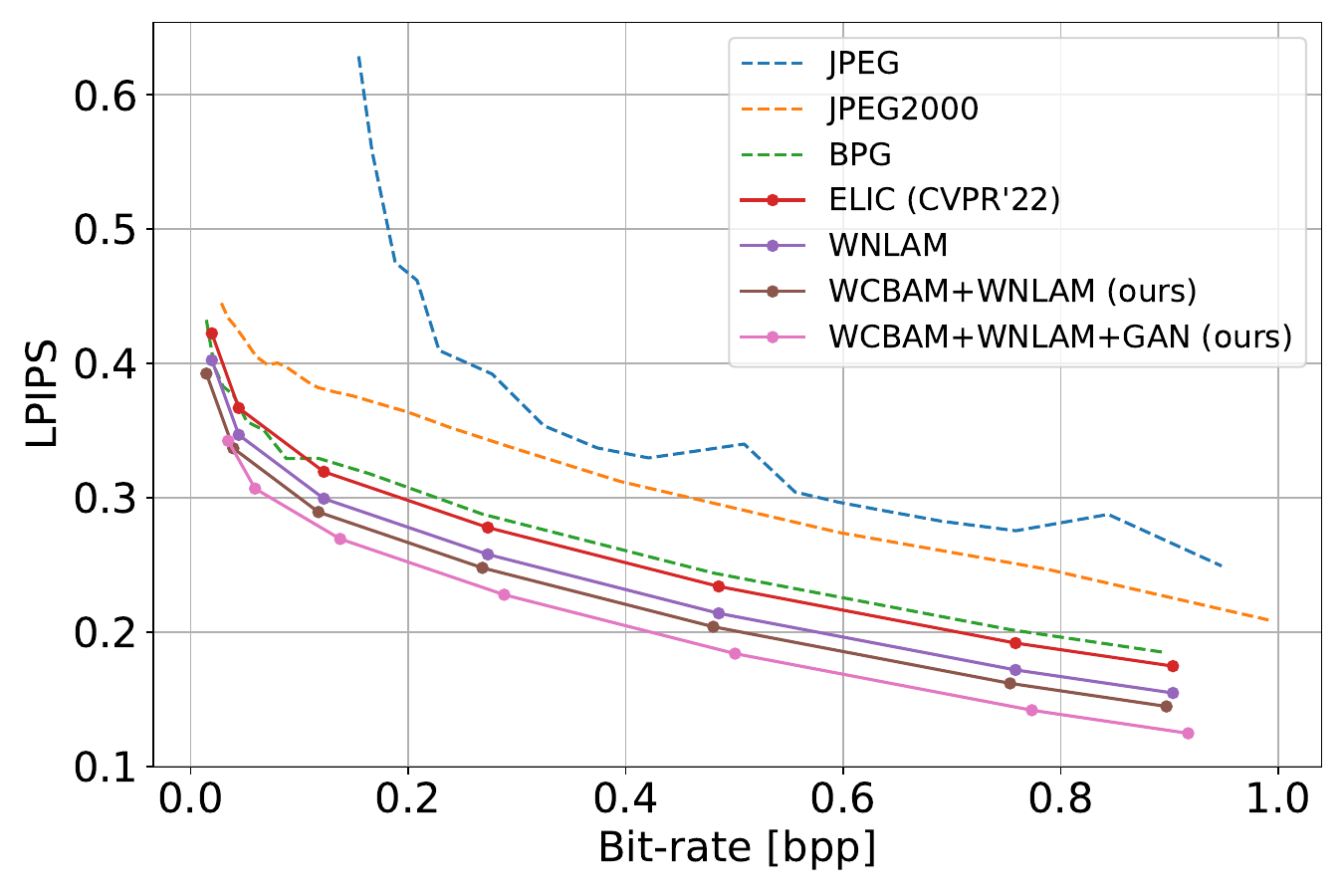}
    \caption{Rate-distortion curve. Distortion is measured by the LPIPS metric (lower is better) as described in Section \ref{sec:compression:objective}. As shown, GAN performance in generating high-quality images can be quantified by this metric.}\label{fig:rd-lpips}
\end{figure}

%Impact of placing the decoder into the generator of a GAN network and scoring the outputs of the generator by the discriminator as shown in \minor{Fig.} \ref{fig:network-arch}. 
As emphasized in Fig. \ref{fig:visual-comparison}, the adversarially trained decoder results in better visual quality of the reconstructed image than the autoencoder only trained with the window-based non-local and convolutional block attention mechanisms. Conventional metrics like PSNR and MS-SSIM are unable to capture the higher perceptual quality of the GAN-reconstructed images. It is empirically shown \citep{zhang2018lpips}  that LPIPS can show the merit of an adversarially trained network. LPIPS is known as a measure of similarity between an image and its reconstruction and it is shown that it is consistent with the human judgment of the quality of reconstructed images \citep{zhang2018lpips}. %Values correspond to the human judgment of the quality of reconstructed images. 
As shown in Fig. \ref{fig:rd-lpips}, the adversarially trained network performs better than the others which are not trained using GANs. This figure shows that if the human judgment has priority over the PSNR/MS-SSIM, training adversarially is the best option.
%to be chosen.
% Add the ACWE File

% This section can be added to a project using the command \input{ACWE} This section relies on the figures in the \fig\ACWE folder, and the citations added to the .bib file that are marked "% Added for ACWE"
% \renewcommand{\revision}[1]{{\color{Sepia}#1}}

\subsection{Example of the Impact on Downstream Use Applications} \label{sec:experiments:ACWE}

\begin{figure*}
    \centering
    \includegraphics[width=0.78\linewidth]{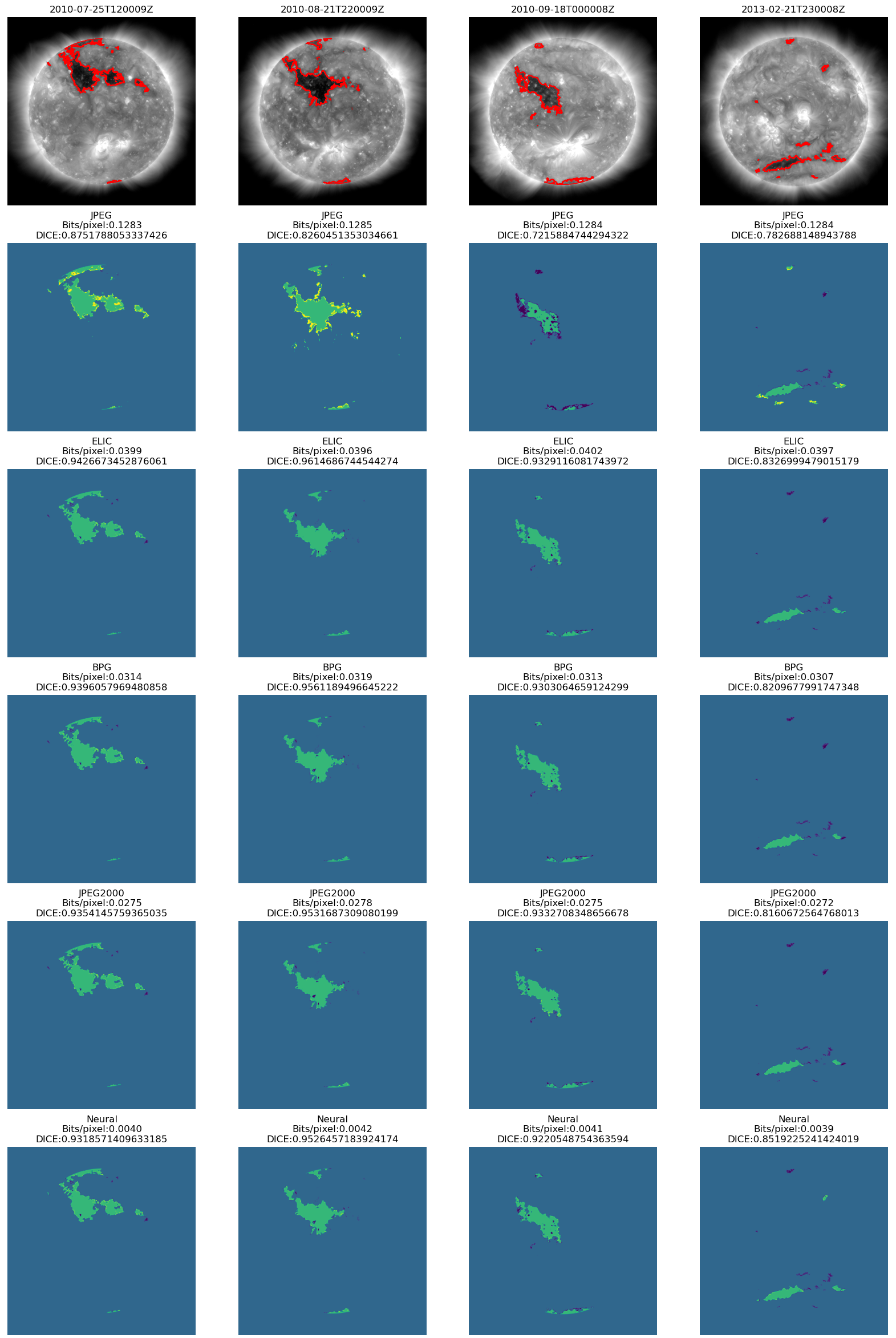}
    \caption{Effects of proposed compression scheme on CH detection via ACWE. The top row images are the original solar EUV images. At this wavelength CHs appear as dark regions. The CH segmentation on the original images is shown in the same row. The remaining images are comparisons of the original ACWE segmentation with the segmentation generated from the compressed images with different compression schemes. Last row contains the results of segmentation on reconstructed images through our proposed neural codec at sub $\sim0.01$ bitrates, a bitrate regime not achievable by other codecs preserving the same DICE coefficient. In each image purple regions were only identified as CHs in the original segmentation, yellow regions were identified as belonging to a CH only in the compressed image segmentation, green regions were identified as CHs in both segmentations, and blue regions were not identified as CHs in either segmentation.}\label{fig:ACWEneuralExample}
\end{figure*}

In order to provide an example of the effects of the proposed compression scheme on data science applications, the coronal hole (CH) detection and segmentation scheme outlined in \cite{boucheron2016segmentation}, which is an extension of the active contours without edges (ACWE) algorithm of \cite{chan2001}, was applied to four 193~{\AA} Solar EUV images. The effects of compression were determined by comparing the similarity of the resulting CH regions at various bit rates to the CH regions identified on the original Level 1 193~{\AA} EUV images. 

CHs are regions of low-density, low-temperature plasma within the Sun's corona. These regions are associated with open magnetic field lines, and are sources of high-speed solar wind \cite{Altschuler1972,Munro1972,Wang1990,Wang1996}. For this reason, accurate delineation of CH boundaries and extents is important for accurate space weather modeling and prediction \cite{Wang1990,Arge2003,Arge2004}. The segmentation scheme outlined in \cite{boucheron2016segmentation} combines threshold-based detection, which is a common method of detecting CHs \cite{Reiss2021}, with a region refinement algorithm that further refines the initial segmentation based on the homogeneity of the region. By using the algorithm of \cite{boucheron2016segmentation} we can verify that region intensity and underlying structure necessary for accurate delineation of coronal holes are both preserved in the compressed images.

ACWE is an iterative process wherein an initial contour or ‘seed’ is manipulated on a pixel-by-pixel basis across multiple iterations in order to minimize an energy functional. The ACWE energy functional is minimized by balancing three forces which seek to 1: minimize the length of the contour, 2: maximize the homogeneity of the foreground (or CH region), and 3: maximize the homogeneity of the background or non-CH region \cite{chan2001}. Each force is subject to a user-defined weight which defines the relative importance of achieving each goal \cite{chan2001}. Unlike the process outlined in \cite{boucheron2016segmentation}, ACWE was performed on the images at the original resolution of $4096\times4096$ pixels. The rest of the process follows the method outlined in \cite{boucheron2016segmentation}, wherein the images are corrected for limb brightening following the method outlined in \cite{verbeeck2014}. An initial seed is then generated by selecting all on-disk pixels with an intensity $\leq\alpha\times QS$ where $QS$ is the mean intensity of the quiet Sun, and the seeding parameter ($\alpha$) is defined as $\alpha=0.3$. From there ACWE was performed on the on-disk region using the length constraint $\mu=0$, and the ratio of foreground to background homogeneity parameters $\lambda_i/\lambda_o=50$.
\begin{figure*}[tp]
    \centering
    \subfigure[JPEG]{\includegraphics[width=0.49\textwidth]{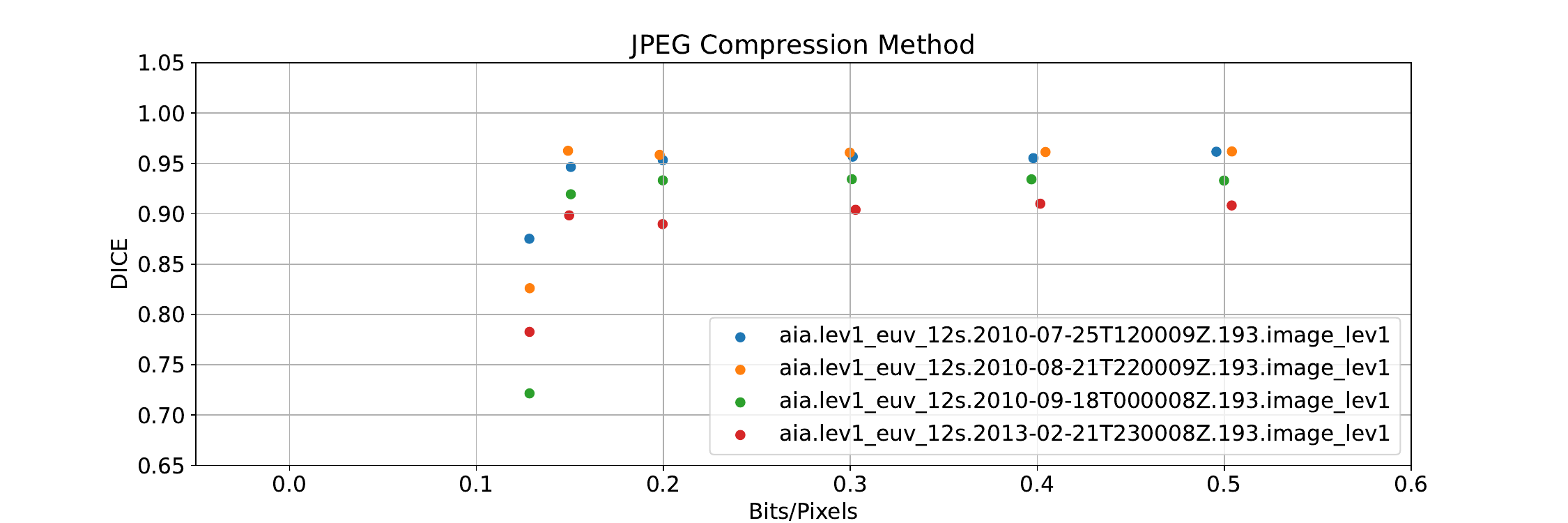} \label{fig:acew-dice-jpeg}}
    \hfill
    \subfigure[JPEG-2000]{\includegraphics[width=0.49\textwidth]{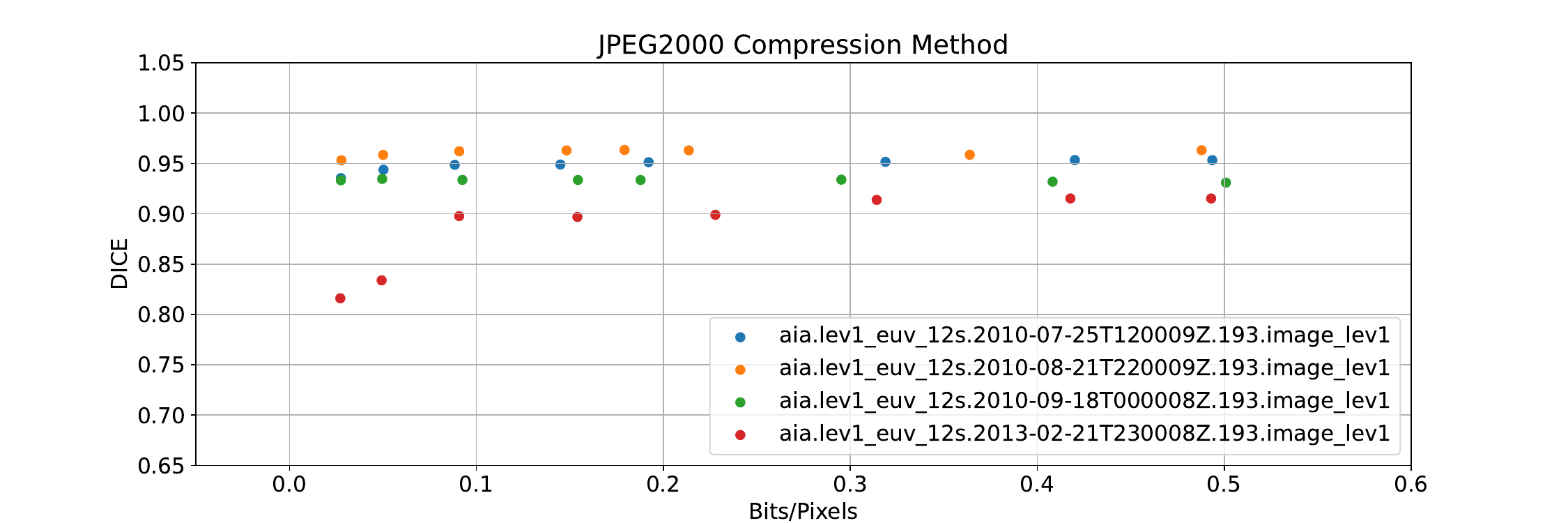}\label{fig:acew-dice-jpeg2000}}
    \vfill
    \subfigure[BPG]{\includegraphics[width=0.49\textwidth]{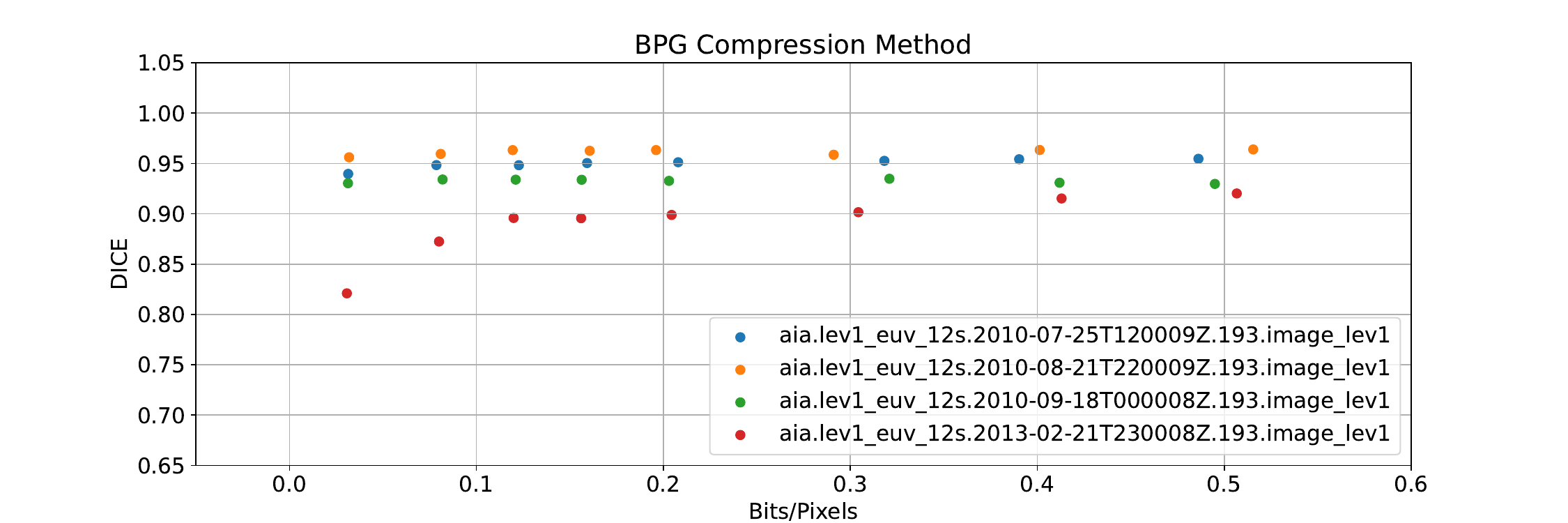} \label{fig:acew-dice-bpg}}
    \hfill
    \subfigure[ELIC]{\includegraphics[width=0.49\textwidth]{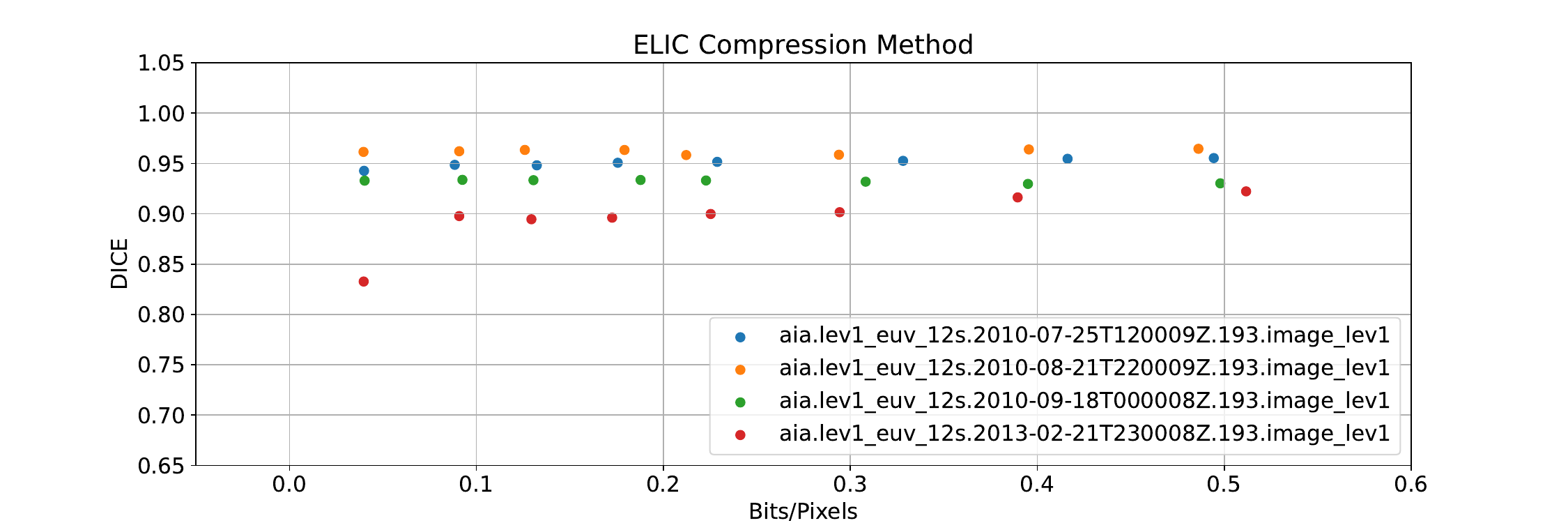}\label{fig:acew-dice-elic}}
    \vfill
    \subfigure[Our Neural Codec]{\includegraphics[width=0.60\textwidth]{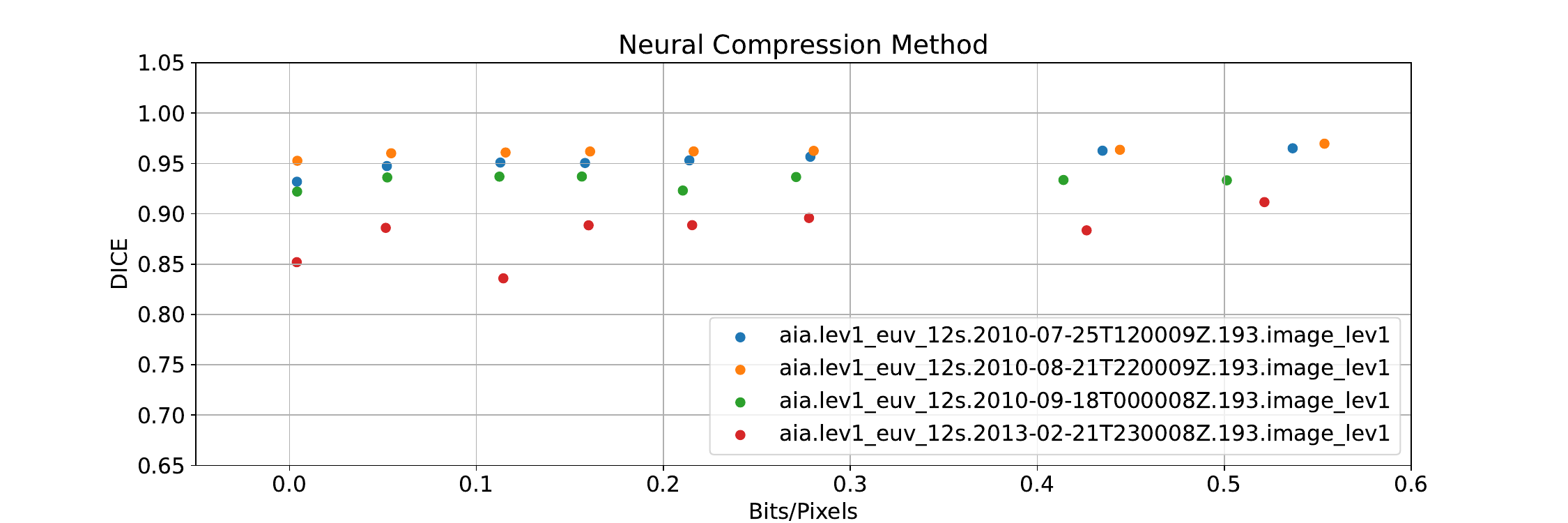} \label{fig:acew-dice-neural}}
    \caption{Effects of different compression schemes on CH detection via ACWE, expressed as DICE coefficient as a function of bit-rate.}
    \label{fig:ACWE-full}
\end{figure*}
The compression process was evaluated using four 193~{\AA} Solar EUV images with record times (as expressed by the \verb.T_REC. keyword) 2010-07-25 T12:00:02Z, 2010-08-21 T22:00:02Z, 2010-09-18 T00:00:02Z, and 2013-02-21 T23:00:01Z. In order to prepare the images for the compression process, the original Level 1 EUV image (at their native spatial resolution of $4096\times4096$ pixels) were clipped to the intensity range of $[20,2500]$. Once clipped, a $\log_{10}$ transform was applied to the intensity levels within images. The resulting intensities where then mapped to 255 discrete levels before performing the compression using the proposed network (with both WNLAM and WCBAM attention). For this process eight models were developed using the hyper-parameters $\lambda=\{0.0005, 0.0035, 0.007, 0.009, 0.011, 0.0125, 0.019, 0.025\}$. Once a compressed image for each EUV input was generated from each model, the images were then restored by reversing the intensity mapping and reversing the $\log_{10}$ transform prior to performing ACWE.

An example of the effects of the proposed compression scheme on the final CH segmentation is presented in Fig. \ref{fig:ACWEneuralExample}.
In this figure the output of compressed images by five different compression algorithms are compared, by showing the original image and its correct segmentation on the first row for four different images of the Sun. The subsequent rows are the highest compression rate achievable by the mentioned algorithm and as it can be seen the proposed neural compression method can still deliver in the sub-0.01 bpp regime.
Each image in Fig. \ref{fig:ACWEneuralExample} is a comparison between the two segmentations (original and compressed) wherein purple regions were only identified as CHs in the original segmentation, yellow regions were identified as belonging to a CH only in the compressed image segmentation, green regions were identified as CHs in both segmentations, and blue regions were not identified as CHs in either segmentation. Within this image set, and the remaining cases tested, the discrepancies between segmentations are generally limited to small-scale structures, usually along the boundary of CH regions. This suggests that the proposed compression scheme is able to preserve the overall structure (and intensity) of coronal hole regions within the solar EUV image. 

Across the four images, the effects of compression were evaluated by computing the DICE coefficient, defined as 
\begin{equation}
    DSC = \frac{2|S_1\cap S_2|}{|S_1|+|S_2|},
\end{equation}
between the CH segmentation generated by ACWE from the original EUV image ($S_1$), and the segmentation generated from the compressed image ($S_2$), where $|\cdot|$ denotes cardinality and $\cap$ denotes intersection. These results, which are shown as a function of bit rate of the compressed image in Fig. \ref{fig:ACWE-full},  corroborates the observations seen in Fig. \ref{fig:ACWEneuralExample} by showing a high similarity between segmentations across all bit rates and more desirably in the extremely compressed regimes.

It should be noted that applying a log transformation to the intensities within an image increases the number of output intensity levels used to represent low intensities while decreasing the number of output levels used to represent high intensities% \cite{DIPbook}
. For this reason, more of the 255 discrete intensity levels were allocated to preserving fine detail at the low-intensity range of the image, which, in turn, improved the quality of the ACWE segmentation \cite{grajeda2023}. 

To help disentangle the effects of this advantage, this same prepossessing was applied to the same EUV images before compressing the images using other compression, then reversing the process in the same manner. The DICE coefficient (again, compared to the segmentation of the original EUV image) as a function of the bit-rate of the compressed image for the other compression schemes is presented in Fig. \ref{fig:ACWE-full}. It should be noted that the other compression schemes begin to show signs of degradation in image features that negatively impact CH segmentation at compression rates of $\sim0.12$ bits per pixel, and cannot produce images with a compression rate $<0.1$ bits per pixel for these 4K by 4K images, unless sacrificing DICE measure. One the other hand, the neural compression scheme proposed here continues to perform well beyond this threshold, suggesting that the neural compression scheme is able to better preserve the features that are relevant to this application.

\section{\textbf{Conclusion}}\label{sec:conclusion}
In this work, we have shown how an effective image compression scheme based on trainable neural networks could be utilized for ad-hoc applications like images from NASA's SDO mission. We explored the effectiveness of window-based spatial and cross-channel attention mechanisms in an adversarially trained neural network to improve the performance of compression in terms of rate-distortion-perception trade-off.  It was shown that neural compression algorithms may be able to benefit data-intensive space missions with minimal degradation in downstream scientific tasks such as coronal hole segmentation as described in this work.

\subsubsection*{Acknowledgments}
This research is based upon work supported by the National Aeronautics and Space Administration (NASA), via award number 80NSSC21M0322 under the title of \emph{Adaptive and Scalable Data Compression for Deep Space Data Transfer Applications using Deep Learning}. 

\bibliography{tmlr}
\bibliographystyle{tmlr}

\end{document}